\documentclass[12pt]{article}
\usepackage[dvips]{graphicx}

\begin{document}
\title
{"Subthreshold" $e^+e^-$ pairs production on photon colliders}
\author{   A.Kochanov, S.Polityko
\\Irkutsk State University}
\maketitle

\begin{abstract}

In this paper we consider the process of "subthreshold"
electron-positron pairs creation in the region of laser
conversion. The total number of positrons and their distribution
are obtained. This phenomena is offered for use as a good test to
examine nonlinear effects of quantum electrodynamics on TESLA.

\end{abstract}

\section{Introduction}
Since the end of the eighties and by this time  works on designing
and development of physical programs for $ \gamma e $ and $ \gamma
\gamma $ - colliders are under development in different countries.
Now there are projects on their creation in USA [1], Germany [2],
Japan [3]. Physical programs for these colliders created as a
result of long-term cooperation of the representatives of many
high-energy physics centres are stated in "Conceptual Design
Reports" [1-3]. In these projects electron-photon and photon -
photon beams are supposed to be obtained on the basis of linear
accelerators with $e^+e^-$ beams.

One of the best methods of obtaining intensive $ \gamma $ beams is
the use of the  Compton backscattering of laser light on an
electron beam of the linear collider. For the first time in works
[4] it was shown, that on the basis of linear colliders with
$e^+e^-$ beams it is possible to realize $e\gamma $ and $\gamma
\gamma$ - beams with approximately the same energies and
luminosities, as for initial electron beams. The necessary
intensive bunches of $\gamma$ quantums were offered for receiving
at scattering of powerful laser flash on electron bunches of these
accelerators.

The small sizes of linear colliders beams make it possible to
obtain conversion coefficient (the attitude of number of
high-energy photons to number of electrons in a bunch)
$k=N_\gamma/N_e\sim 1$ at energy of laser flash in some Joules,
i.e. it is possible to convert the most part of electrons to
photons.

The detailed description of the scheme of an electron beam
conversion in $\gamma$ beam, the basic characteristics of
$e\gamma$ and $\gamma\gamma $ collisions, problems of a background
and calibration of luminosity were considered in detail in [5].

The region of laser conversion $e\to \gamma$ is unique by its
physical properties. It is the region of an intensive
electromagnetic field (the focused laser bunch). This fact allows
one to investigate such processes of nonlinear quantum
electrodynamics as radiation of a photon by electron in a field of
an intensive electromagnetic wave, and also "subthreshold" pairs
production \cite{NQED}.

At sufficient power of laser flash in the field of conversion the
processes are essential due to absorption from a wave more than
one of laser photons simultaneously
\begin{equation}
e^-(p)+n\gamma_0(k_0)\to e^-(p^\prime)+\gamma(k),\quad n\ge1;
\label{1}
\end{equation}
\begin{equation}
\gamma(k)+s\gamma_0(k_0)\to e^-(p_+)+e^+(p_-),\quad s\ge1;
\label{2}
\end{equation}

Processes (\ref{1}),(\ref{2}) represent nonlinear by intensity of
a field processes of interaction electrons and photons with a
field of an electromagnetic wave. The first of these nonlinear
processes results in expansion of spectra of high-energy photons
and occurrence of additional peaks in spectra of scattered
radiation due to absorption of several photons from a wave, and
the second one effectively reduces a threshold of $e^+e^-$ pairs
creation. The interaction of electrons and positrons with a field
of an electromagnetic wave results in effective increase of their
masses\footnote{Here we use the system $\hbar =c=1$}:
\begin{equation}
m^2 \to m^2(1+\xi^2), \label{3}
\end{equation}
which is characterized by parameter of intensity of a laser wave
$\xi^2$:
\begin{equation}
\xi^2 = n_\gamma \bigg({4\pi \alpha \over m^2\omega_0}\bigg )=
-{e^2a^2\over m^2}, \label{4}
\end{equation}
where  $n_\gamma $- density of photons in a laser wave, $\omega_0$
- their energy, $a$ - amplitude of classical 4-potential of
electromagnetic wave, $e$ - a charge of electron. Regular research
of nonlinear Breit-Wheeler (\ref{2}) and Compton (\ref{1})
processes was carried out in \cite{Ritus},\cite{LD}.

Now the  area of nonlinear effects is rather actual and is of
great interest because here essential are the processes of
radiation due to absorption from a wave of a few of photons, and
their probabilities are essentially nonlinear functions of
intensity of a field. Recently on accelerator SLAC \cite{SLAC} a
series of experiments E-144 with check of predictions  of
nonlinear QED was finished in the field of parameter $\xi \sim 1$
that became possible due to use of the supershort and rigidly
focused laser pulses. Thus for the first time the experiment was
set up in which the process of $e^+e^-$ - pair production at
participation of only real, instead of virtual photons was carried
out.

\section{Spectra of high-energy photons}

The main features of the conversion are described by a quantity
$x$ which is determined via the initial electron beam energy $E_e$
and the laser photon energy $\omega _0$ as $$x={4E_e\omega_0\over
m^2}.$$
 The differential
probability of process of radiation of a photon by electron
performed by a summation over polarizations of final electron and
a photon has the following form \cite{Ritus}:

\begin{equation}
{dW\over dy}={\pi \alpha^2\over 2xm^2}\sum_{n=1}^\infty
(F_{00}+2\lambda P_cF_{11}), \label{5}
\end{equation}
 $$ F_{00}=-4{J_n(z)^2\over
\xi^2}+(2+uy)[J_{n-1}^2(z)+J_{n+1}^2(z)-2J_n^2(z)], $$ $$
F_{11}={(2+u)u(1-2r_n)\over u+1}[J_{n-1}^2(z)-J_{n+1}^2(z)], $$ $$
u={y\over 1-y},\quad z={2n\xi \over \sqrt{1+\xi^2}}\sqrt{{u\over
u_n} (1-{u\over u_n})}, \quad u_n={nx\over 1+\xi^2},\quad$$ $$r_n=
{y\over nx(1+\xi^2)(1-y)}, $$ $J_n(z)$ is the  Bessel functions of
nth order, $y=E_\gamma /E_e, E_\gamma $ is energy of high-energy
photon.

The expression  in the sum  (\ref{5}), determines probability of
radiation  of n- harmonics by electron in a field of circular
-polarized electromagnetic wave (from a wave n laser photons can
be absorbed ). The change of a variable $y$ corresponds to the
change of a variable $u$: $0<u<u_n$
 $$ 0<y<{nx\over nx+1+\xi^2}. $$
The influence of nonlinear effects results in the fact that the
maximum energy of high-energy photons of the first harmonic
$(n=1,\xi^2 \neq 0)$ decreases in comparison with the maximum
energy of photons in usual Compton effect and the energy of the
$\gamma$ - quanta formed at absorption from a wave of several
photons exceeds energy, achievable in usual Copmton effect.

\begin{figure}
\hspace{1cm}
\includegraphics[bb=110 630 240 770,scale=0.4]{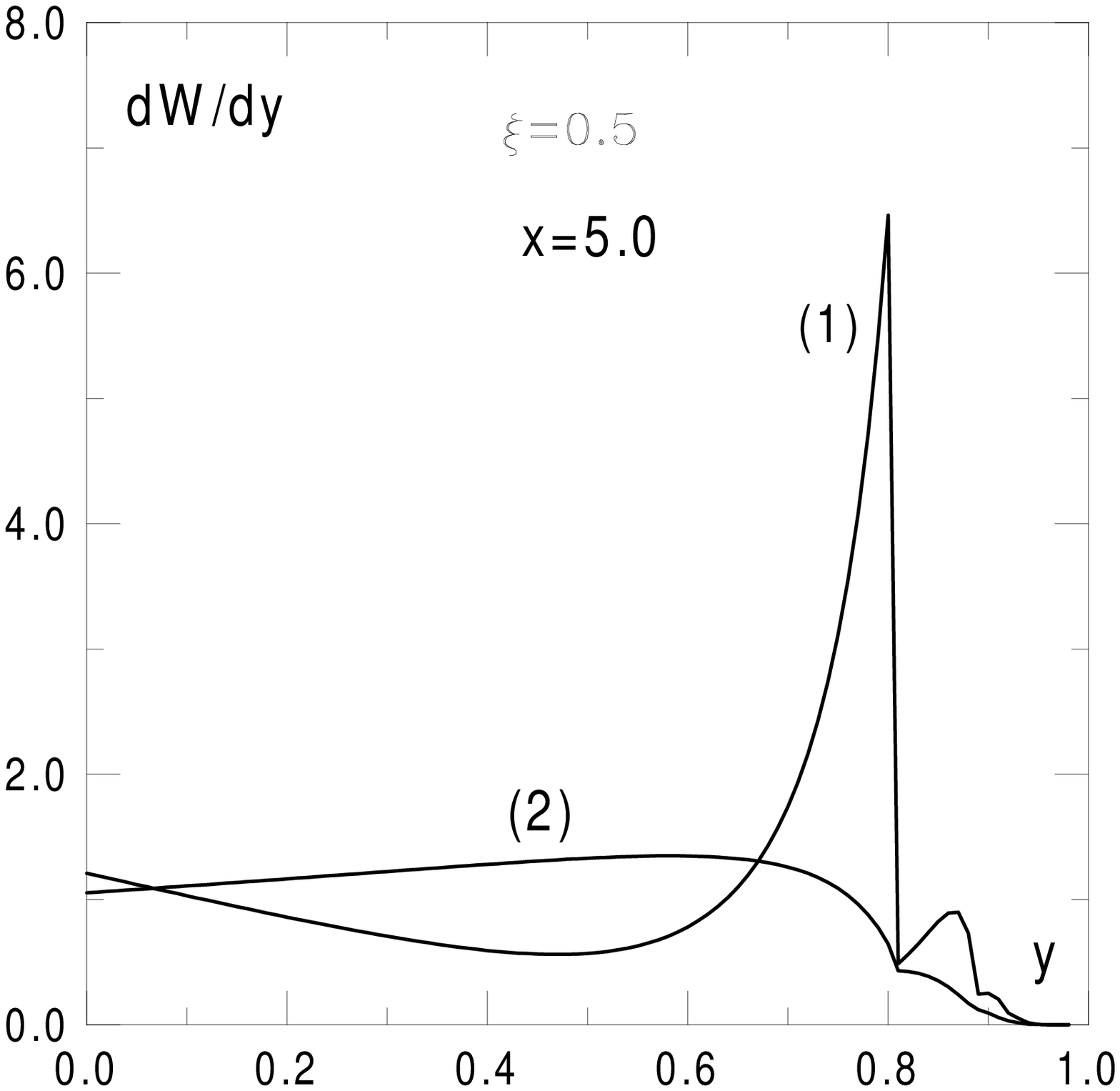}
\hspace{9.5cm}
\includegraphics[bb=340 630 480 770,scale=0.4]{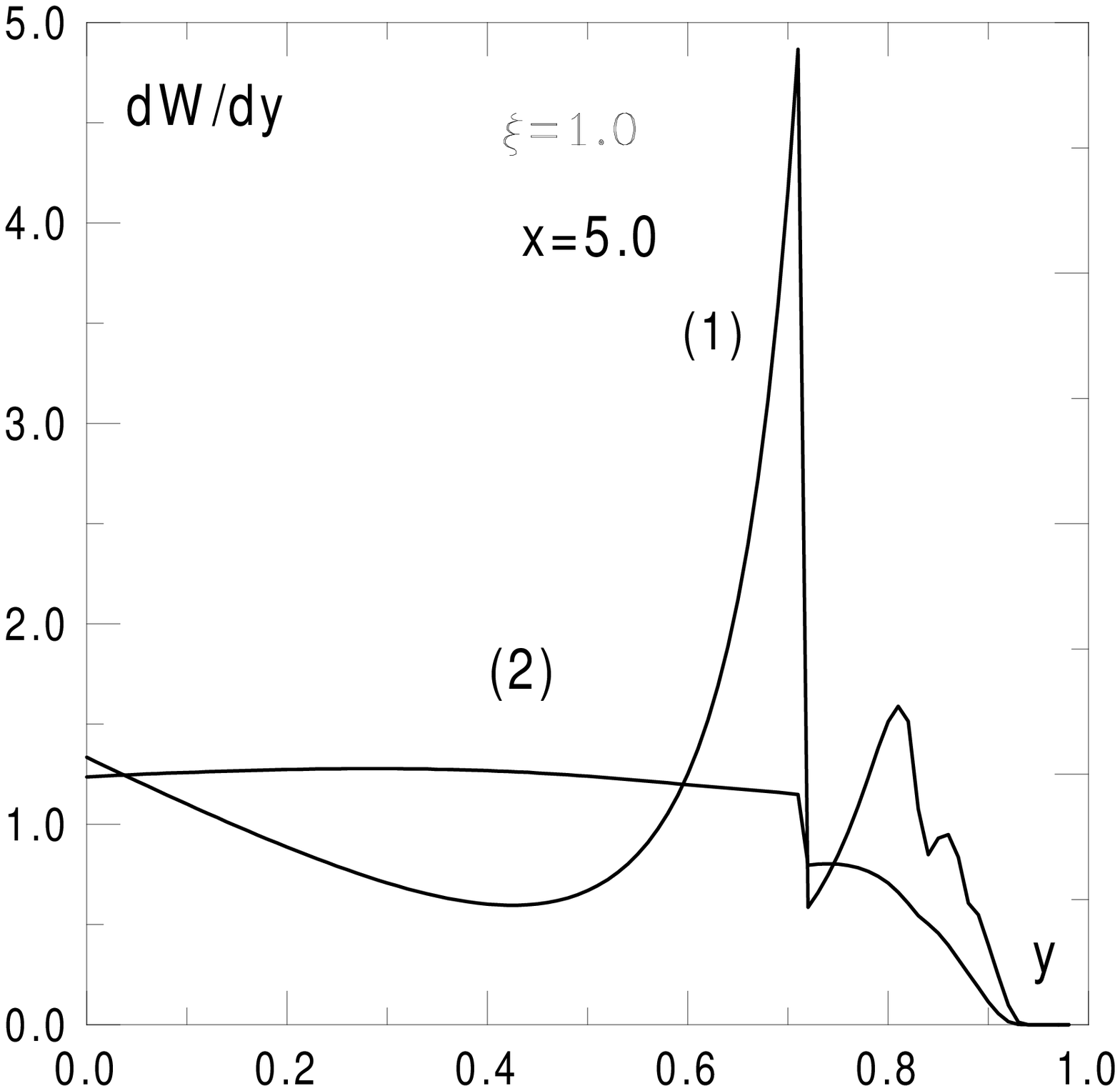}
\vspace{6cm} \caption{ Spectra of a high-energy photon in a
nonlinear case at various values of polarizations of initial
particles: $(1)-2\lambda_e P_c=-1, (2)-2\lambda_e P_c=+1$  and
various values of parameter $\xi =0.5$ , $1.0$}
\end{figure}

Results of numerical calculations of power spectra of photons in a
nonlinear case at $x=5$ are given in Fig 1. Apparently from these
figures, the account of nonlinear effects results in essential
change of spectra in comparison with spectra of usual Copmton
scattering. First, simultaneous absorption from a wave of several
laser photons results in expansion of spectra of rigid $\gamma $
quanta and occurrence of the additional peaks, appropriate to
radiation of harmonics of higher order. This expansion at the same
parameter x increases with the intensity of a wave. Second,
effective electron weighting results in compression of spectra,
i.e.in shift of the first harmonic aside smaller values y. With
increase of parameter x compression of the first harmonic
decreases.

\section {"Subthreshold" $e^+e^-$ pairs creation}

The process discussed in this section, is the good test on check
of nonlinear quantum electrodynamics.

The "hard" photons propagated in the region of laser conversion
can produce $e^+e^-$ pairs in collision with s laser photons
simultaneously $$ \gamma + s\gamma _0 \to e^+ +e^- $$ Threshold
value for energy of a photon is defined from the relation $$
(k+sk_0)^2=4m^2(1+\xi^2) $$ Here $k$ , $k_0$ - 4- momentum of
photons $\gamma $ and $\gamma _0$. The maximum energy of Compton
photon emitted at interaction of electron with n laser photons is
$$ {E_\gamma \over E_e}={4nE_e\omega _0\over 4nE_e\omega
_0+m_*^2}= {nx\over nx+1+\xi^2} $$ Thus the corresponding value of
electron energy for $e^+ e^-$ pairs creation at absorption from a
laser wave s of photons is $$
E_e^{th}={m^2(1+\xi^2)(1+\sqrt{1+s/n})\over 2s\omega_0} $$ In
particular if the "hard" photon is emitted as a result of
interaction with one laser photon$$
E_e^{th}={m^2(1+\xi^2)(1+\sqrt{1+s})\over 2s\omega_0} $$ $$
x_{min}=2(1+\xi^2)(1+\sqrt{1+s}) $$ When $x<2(1+\sqrt{2})\approx
4.8$ electron-positron pairs will be created only due to nonlinear
processes QED. Thus at the first stage of realization of projects
NLC (in particular for TESLA) at x=4.5 occurrence of the big
number of positrons will be a consequence of nonlinear effects
QED.

Calculated in \cite{Ritus} the probability of $e^+e^-$ pair
production at interaction with circle-polarized laser photons can
be submitted by the high-energy non-polarized photon as
\begin{equation}
W_{e^+e^-}={e^2m^2\over 8\pi E_\gamma }\sum_{s=1}^\infty
\int_1^{u_s} dw_s, \label{6}
\end{equation}

$$ dw_s={du\over u\sqrt{u(u-1)}}\biggr[J_s^2(z)+\xi^2(2u-1)
[({s^2\over z^2}-1)J_s^2(z)+J_s^{\prime 2}] \biggl] $$ $$
u_s={sxy\over 4(1+\xi^2)},\quad z={2\xi s\over \sqrt{1+\xi^2
}}{\sqrt{u(u_s-u)}\over u_s} $$

The total number of the produced positrons can be obtained by
averaging on an energy spectrum of Compton photons (\ref{5})
\begin{equation}
N_{e^+e^-}={1\over 4}kN_e\tau {e^2m^2\over 8\pi E_\gamma }
\int_0^{y_{max}}\sum_{s=1}^\infty \int_1^{u_s} dw_s{1\over
W}{dW\over dy}dy \label{7}
\end{equation}
Here $y_{max}=nx/(nx+1+\xi^2)$. Expression (\ref{7}) contains two
factors 1/2. The first one arises at the account of relative
movement of bunches of photons $\gamma $ and $\gamma _0$. The same
multiplier appears because of the fact that  photons interact in
the average with the half of a laser bunch (the length of a laser
pulse does not exceed the sizes of area of conversion and
distribution of density of bunches in a direction of movement is
homogeneous). In Fig. 2 the number of the created positrons per
one electron is shown depending on x at various values of
parameter $\xi $, value of factor of conversion $k=0.5$, the
length of a laser bunch $l\sim 1 cm$.

\begin{figure}
\centering
\includegraphics[bb=170 160 430
650,scale=0.4]{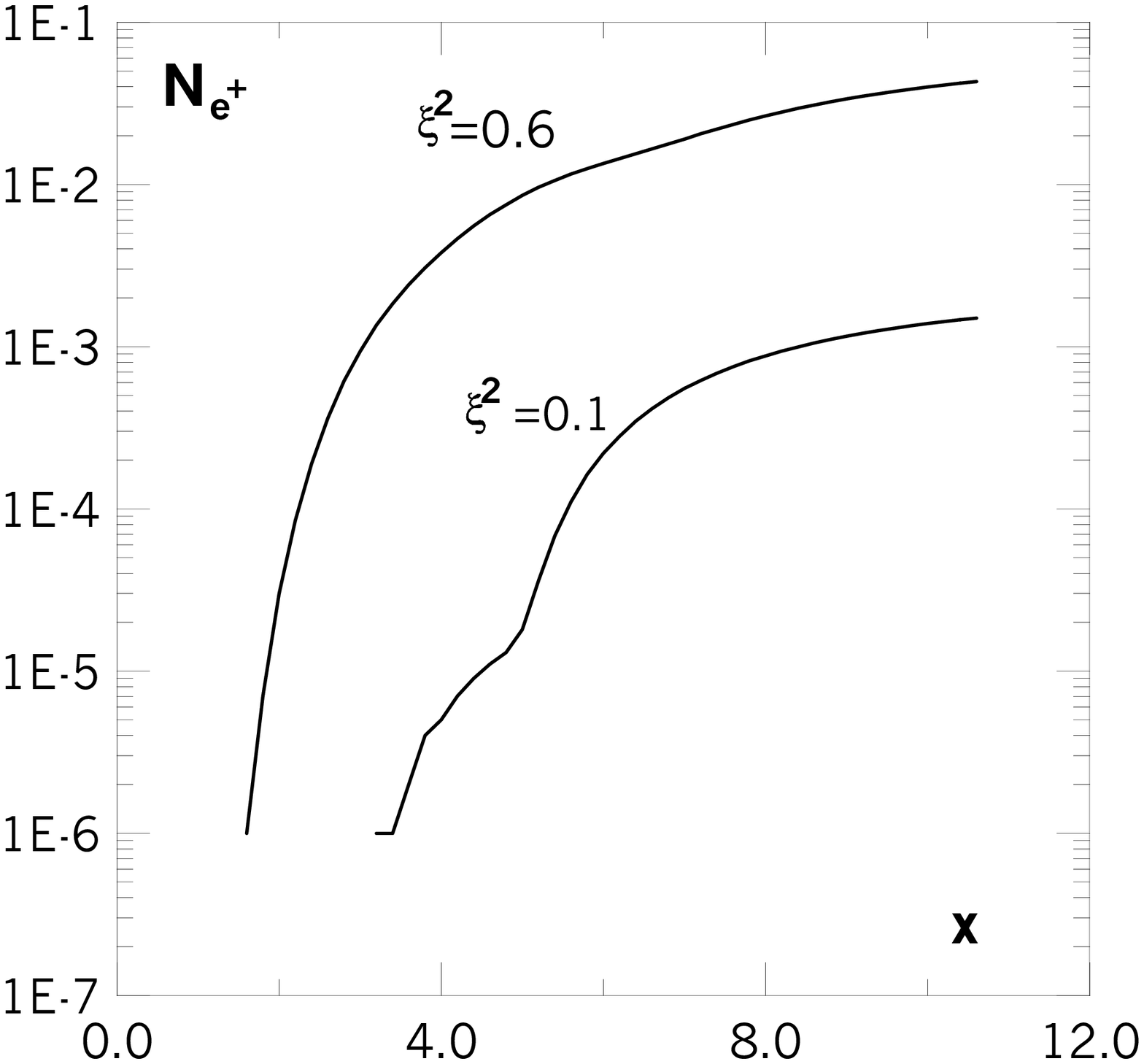}
 \caption{Number of positrons (electrons) created by one initial
 electron at various values of parameter
$\xi $}
\end{figure}

At the first stage of project TESLA will be observed $10^7$ pairs
per one collision. These pairs will form the essential background,
therefore the produced positrons should be removed from the region
of interaction with the help of a magnetic field. To observe the
process of creation of pairs it is necessary to extract the
positrons moving  practically along the  direction of initial
electrons.

\section  { The distribution of created positrons}

We shall consider distribution on energy of the created positrons.
Energies of positrons (electrons) are distributed in the interval

\begin{equation}
\epsilon_\pm ={1\over E_\gamma }(m_*^2u_s-\xi^2m^2u) (1\mp
\sqrt{1-1/u})\pm {1\over 2}E_\gamma (1\pm \sqrt{1-1/u}) \label{8}
\end{equation}
From here it is easy to get restricts of distribution of positrons
on energy $$ \epsilon _{min} \leq \epsilon _+ \leq \epsilon
_{max}, $$ $$ \epsilon _{max} \approx {1\over 2}E_{\gamma
max}(1+\sqrt{1-1/u_s}), $$ $$ \epsilon _{min} \approx {1\over
2}E_{\gamma max}(1-\sqrt{1-1/u_s}). $$

\vspace{1cm} \centerline{Table }
\begin{center}
\begin{tabular}{|c|c|c|c|c|c|} \hline
\multicolumn{3}{|c|} {0.25 \quad TeV}&
   \multicolumn{3}{|c|} {1 \quad TeV}\\  \hline
 $s$& $ \epsilon _{+min}$ \quad (GeV) & $ \epsilon _{+max}$ \quad (GeV)
&
 $s$& $ \epsilon _{+min}$ \quad (GeV) & $ \epsilon _{+max}$ \quad (GeV)
  \\  \hline
  1 &   --  &  --   &       1 &  99.7 & 818.7  \\
  2 &  63.4 & 148.9 &       2 &  46.7 & 910.8  \\
  3 &  35.2 & 188.3 &       3 &  30.8 & 940.5  \\
  4 &  24.9 & 204.7 &       4 &  22.8 & 955.8  \\
  5 &  19.4 & 214.1 &       5 &  18.4 & 964.4  \\
\hline
\end{tabular}
\end{center}
\vspace{1cm}

This table represents borders of positrons energies for various
number s of the absorbed laser photons, parameter $\xi^2 =0.6$.
Differential distribution of the produced pairs on energy of
positrons looks like
\begin{equation}
{dw_{e^+e^-}^s\over d\rho }={e^2m^2\over 2\pi E_\gamma }
\biggr[J_s^2(z)+\xi^2(2u-1) [({s^2\over
z^2}-1)J_s^2(z)+J_s^{\prime 2}] \biggl], \label{9}
\end{equation}
$$ \rho={\epsilon_+\over E_{\gamma max}},\quad u\approx {1\over
1-(2\rho-1)^2} $$ At small $\xi $ distribution on energy (\ref{9})
becomes $$ {dw_{e^+e^-}^s\over d\rho }  ={\alpha m^2 \over 2Ey}
\xi ^{2s} {s^{2s}\over 2(s!)^2} \left (2{u^{s-1}(u_s-u)^s \over
u_s^{2s}\sqrt{u(u-1)}}+
 {(2u-1)u^{s-2}(u_s-u)^{s-1} \over u_s^{2(s-1)}\sqrt{u(u-1)}}\right )
$$ In Fig. 3a differential distributions of the created positrons
by a "hard" photon of the maximal energy for various number s of
the absorbed laser photons are submitted. Electron energy $E_e=250
GeV, \xi^2 =0.6$. In Fig.3b we show the same distributions, but
averaged in the spectrum of high-energy photons. In Fig. 4a and 4b
distributions for energy of electrons $E_e=1 TeV$ are shown.

\begin{figure}
\hspace{1cm}
\includegraphics[bb=80 530 240 770,scale=0.4]{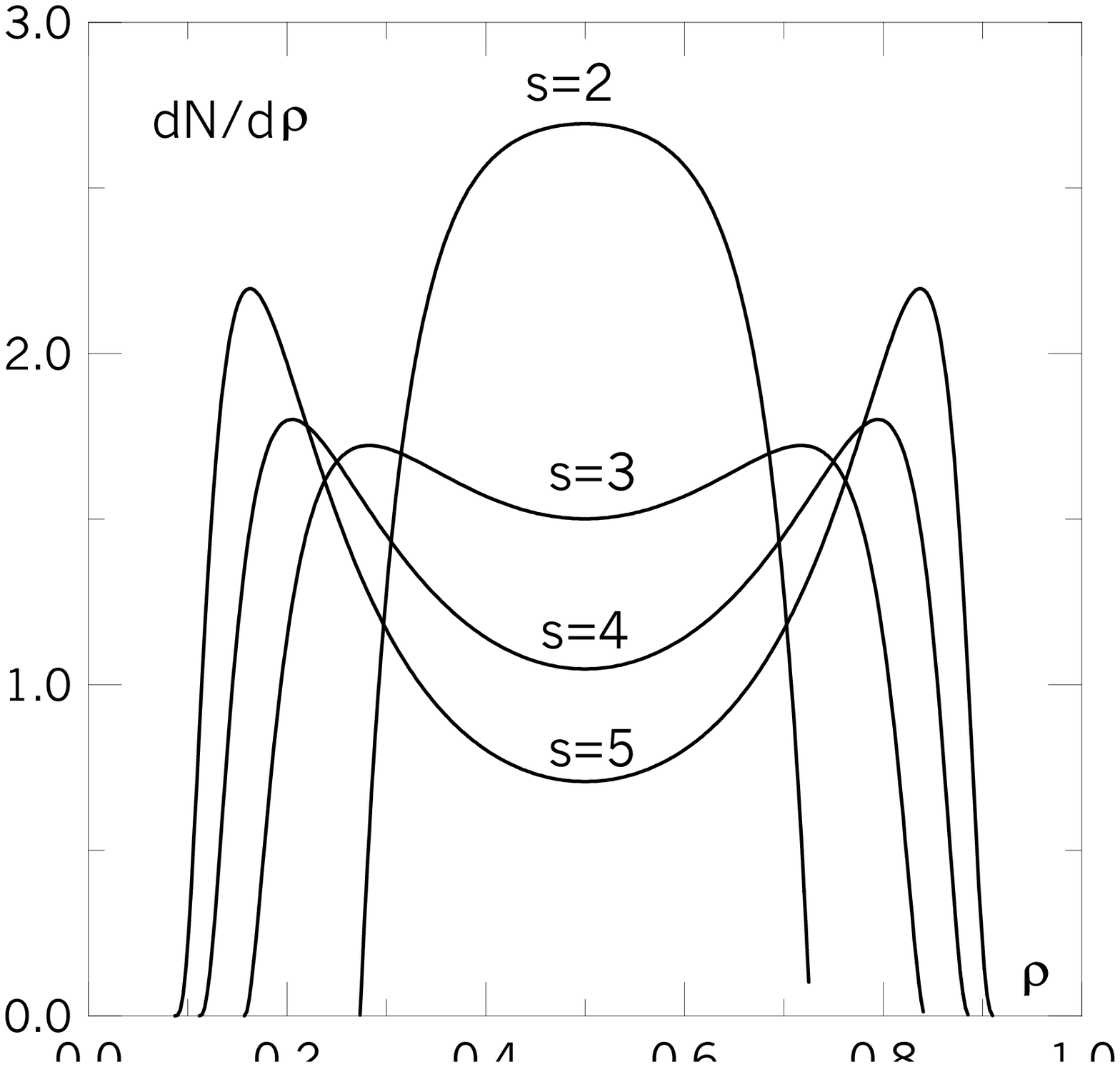}
\hspace{9.5cm}
\includegraphics[bb=340 530 480 770,scale=0.4]{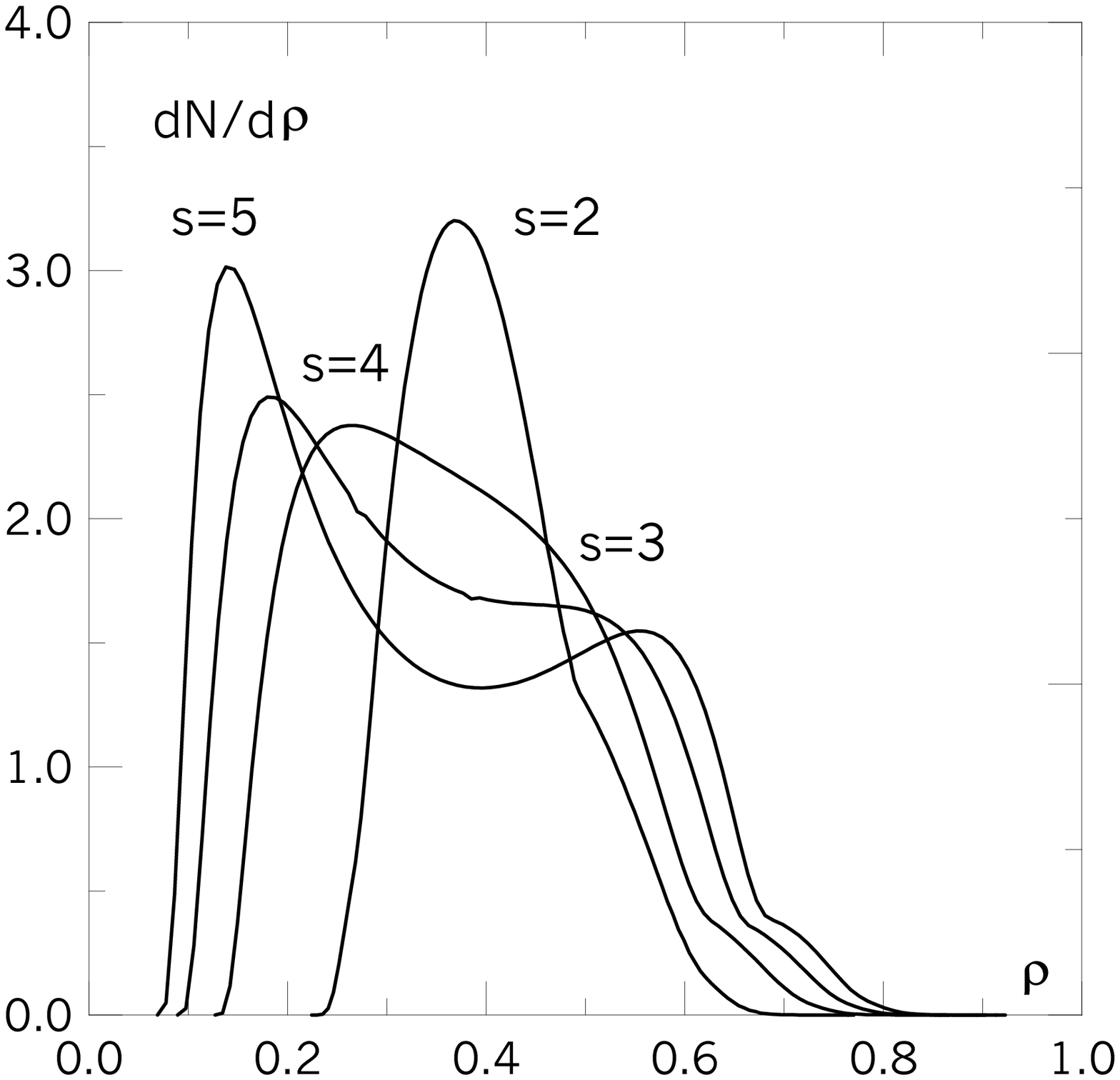}
\vspace{6cm} \caption{ Differential distribution positrons on
energy at  $x=4.5,\xi^2=0.6$  for different number of absorbed
laser photons, (a) - without averaging on a spectrum, (b) - with
averaging on a spectrum.
 }
\end{figure}

\begin{figure}
\hspace{1cm}
\includegraphics[bb=80 530 240 770,scale=0.4]{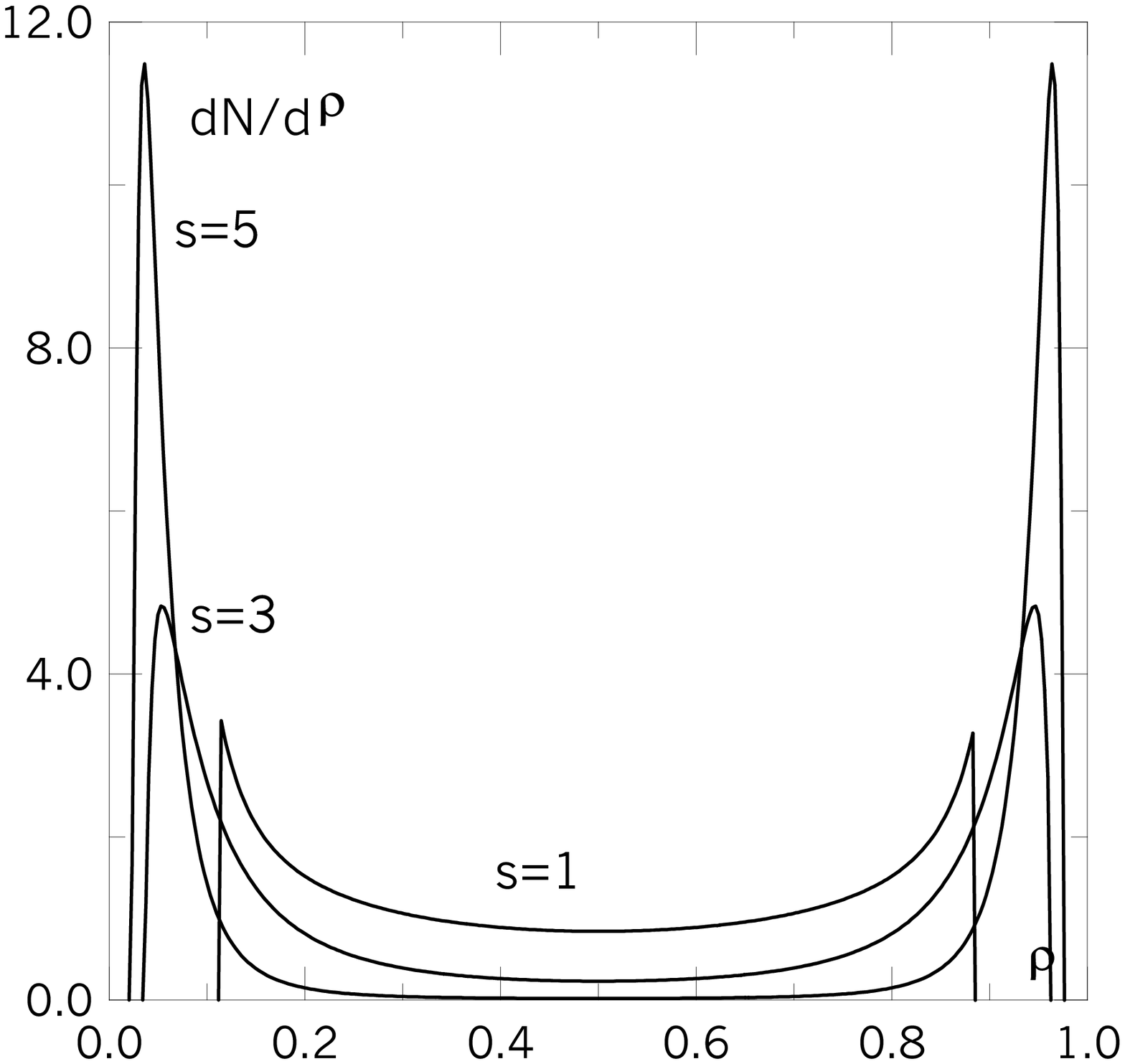}
\hspace{9.5cm}
\includegraphics[bb=340 530 480 770,scale=0.4]{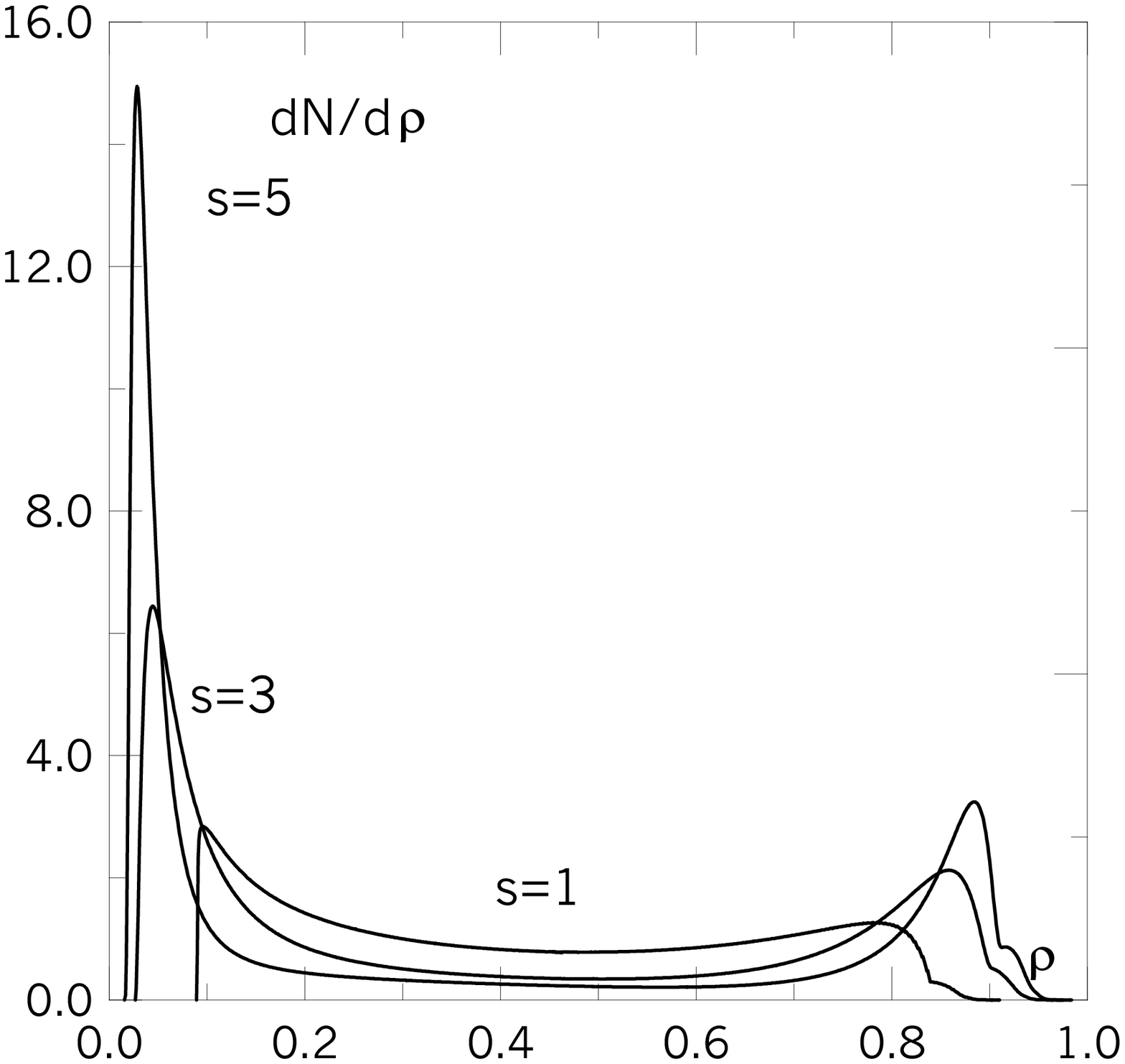}
\vspace{6cm} \caption{Differential distribution positrons on
energy for $x=18,\xi^2=0.6$ for different number of absorbed laser
photons, (a) - without averaging on a spectrum, (b) - with
averaging on a spectrum.
 }
\end{figure}
The total distribution of positrons is given in Fig.5 From Figures
it is possible to see that the maximum in distribution of
positrons after averaging on a spectrum of high-energy photons is
shifted to the bottom border of a spectrum. It enables precise
identification of "subthreshold" $e^+e^-$ pairs creation as effect
of nonlinear quantum electrodynamics. At $x>4.8$ it is possible to
observe "subthreshold" creation of positrons in a positron
spectrum near the bottom border. For energy of initial electron
$E_e=250 GeV$ positrons with energy in an interval $19.4-24.9 GeV$
are created by absorption of 5 laser photons simultaneously. For
energy of initial electron $E_e=1 TeV$, positrons with energies
$\epsilon_+ < 99.7 GeV$ are created only due to nonlinear effects
QED.

\begin{figure}
\centering\includegraphics[bb=170 160 430 650,
scale=0.6]{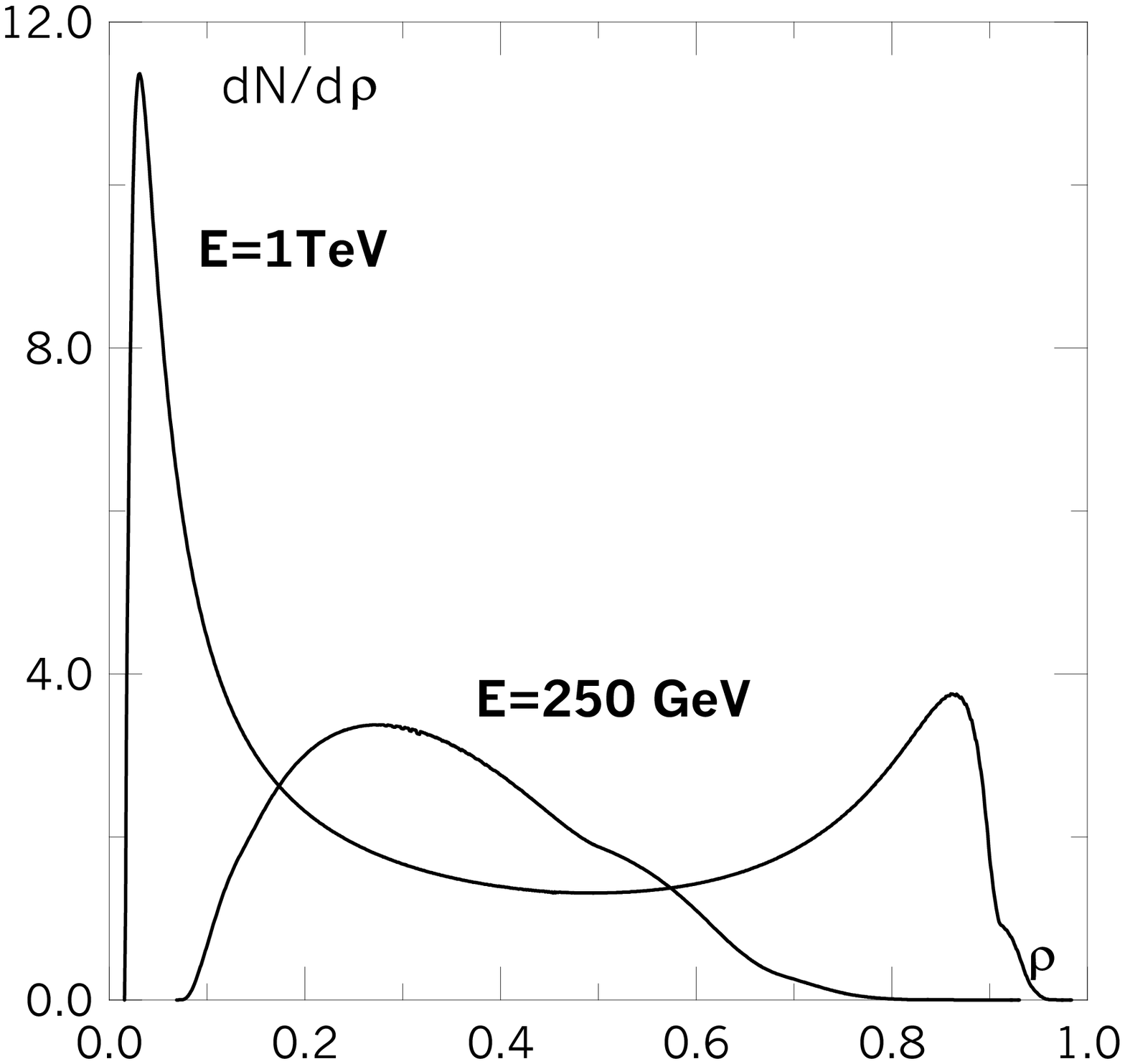}
 \caption{The total distribution of produced positrons for
 different values of electron energy $E_e $}
\end{figure}

Let's note that the contribution of other mechanisms to creation
of $e^+e^- $ pairs is much less. For the Bethe-Heitler process $$
e+s\gamma _0 \to e^++e^-+e $$ threshold values of energy are much
greater $$ E_{th}={2m_*^2\over s\omega_ 0 }, $$ Except for it, the
probability of creation contains the additional factor $\alpha $.

\section{ Acknowledgments}

The authors would like to thank I.F.Ginzburg, G.L.Kotkin and
V.G.Serbo for useful discussions.


\begin{thebibliography}{99}

\bibitem{NLC}
{\it Zeroth-Order Design Report for the Next Linear Collider}
LBNL-PUB-5424,\\ SLAC Report 474, May 1996.

\bibitem{Tesla}
{\it Conceptual Design of a 500 GeV Electron Positron Linear
Collider with Integrated X-Ray Laser Facility}\\ DESY 97-048,
ECFA-97-182; Brinkmann R et al.

\bibitem{JLC}
{\it JLC Design Study} KEK-REP-97-1, April 1997; Watanabe I et al.

\bibitem{GKST}
{\it Ginzburg I, Kotkin G, Serbo V, Telnov V} Pizma ZhETF,
1981,{\bf 4}, 514;\\ Nucl.  Instr.  and  Methods (NIM) 1983 {\bf
205}, 47;\\ {\it Ginzburg I F,  Kotkin G L,  Panfil S L, Serbo V
G,  Telnov V I}  NIM, 1984 {\bf 219}, 5.

\bibitem{Tel1}
{\it Telnov V I} Nucl.Instr. and  Methods {\bf 24} (1990),72\\
{\it Ginzburg I F} Nucl.Instr. and Methods {\bf A 355} (1995), 63

\bibitem{NQED}
{\it Baier V N, Yokoya K} Particle Accelerators, 1994,
Vol.44,pp.77 \\ {\it Ginzburg I F, Kotkin G L, Polityko S I } Sov.
Yad. Fiz., 1984, v.40, 1495\\ {\it Ginzburg I F, Kotkin G L,
Polityko S I } Sov. Yad. Fiz., 1993, v.56, 65

\bibitem{Ritus}
{\it Ritus V I} Trudy FIAN,v.111,1979 [in Russian].

\bibitem{LD}
{\it Berestetskii V, Lifshitz E and Pitaevskii L } Quantum
Electrodynamic, Pergamon press, Oxford, 1982.

\bibitem{SLAC}
 Bula C et all., Phys. Rev. Lett. 76 (1996) 3116.
\end{thebibliography}
\end{document}